\documentclass[aps,prl,preprint,superscriptaddress]{revtex4-2}

\usepackage{graphicx}
\usepackage[T1]{fontenc}
\usepackage[hidelinks]{hyperref}
\usepackage{bm}
\usepackage{color}
\usepackage{xfrac}

\usepackage{placeins}
\usepackage{amsmath}

\makeatletter
\def\maketitle{
\@author@finish
\title@column\titleblock@produce
\suppressfloats[t]}
\makeatother

\begin{document}

\title{Spontaneous anomalous Hall effect arising from an unconventional compensated magnetic phase in a semiconductor}

\author{R.~D.~Gonzalez Betancourt}
\altaffiliation{R.D.G.B. and J.Z. contributed equally to this work}
\affiliation{Institute of Solid State and Materials Physics, Technical University Dresden, 01062 Dresden, Germany}
\affiliation{Institute of Physics, Academy of Sciences of the Czech Republic, Cukrovarnick\'a 10, 162 00 Praha 6, Czech Republic}
\affiliation{Leibniz Institute of Solid State and Materials Research (IFW Dresden), Helmholtzstr. 20, 01069 Dresden, Germany}

\author{J.~Zub\'a\v{c}}
\altaffiliation{R.D.G.B. and J.Z. contributed equally to this work}
\affiliation{Institute of Physics, Academy of Sciences of the Czech Republic, Cukrovarnick\'a 10, 162 00 Praha 6, Czech Republic}
\affiliation{Charles University, Faculty of Mathematics and Physics, Ke Karlovu 3, 121 16 Prague 2, Czech Republic}

\author{R.~Gonzalez-Hernandez}
\email[]{rhernandezj@uninorte.edu.co}
\affiliation{Departamento de Fisica y Geociencias, Universidad del Norte, Barranquilla 080020, Colombia}

\author{K.~Geishendorf}
\affiliation{Institute of Physics, Academy of Sciences of the Czech Republic, Cukrovarnick\'a 10, 162 00 Praha 6, Czech Republic}

\author{Z.~\v{S}ob\'a\v{n}}
\affiliation{Institute of Physics, Academy of Sciences of the Czech Republic, Cukrovarnick\'a 10, 162 00 Praha 6, Czech Republic}

\author{G.~Springholz}
\affiliation{Institute of Semiconductor and Solid State Physics, Johannes Kepler University Linz, Altenbergerstr. 69, 4040 Linz, Austria}

\author{K.~Olejn\'ik}
\affiliation{Institute of Physics, Academy of Sciences of the Czech Republic, Cukrovarnick\'a 10, 162 00 Praha 6, Czech Republic}

\author{L.~\v{S}mejkal}
\affiliation{Institut f\"{u}r Physik, Johannes Gutenberg Universit\"{a}t Mainz, 55128 Mainz, Germany}
\affiliation{Institute of Physics, Academy of Sciences of the Czech Republic, Cukrovarnick\'a 10, 162 00 Praha 6, Czech Republic}

\author{J.~Sinova}
\affiliation{Institut f\"{u}r Physik, Johannes Gutenberg Universit\"{a}t Mainz, 55128 Mainz, Germany}
\affiliation{Institute of Physics, Academy of Sciences of the Czech Republic, Cukrovarnick\'a 10, 162 00 Praha 6, Czech Republic}

\author{T.~Jungwirth}
\affiliation{Institute of Physics, Academy of Sciences of the Czech Republic, Cukrovarnick\'a 10, 162 00 Praha 6, Czech Republic}
\affiliation{School of Physics and Astronomy, University of Nottingham, Nottingham NG7 2RD, United Kingdom}

\author{S.~T.~B.~Goennenwein}
\affiliation{Department of Physics, University of Konstanz, 78457 Konstanz, Germany}
\affiliation{Institute of Solid State and Materials Physics, Technical University Dresden, 01062 Dresden, Germany}

\author{A.~Thomas}
\affiliation{Institute of Solid State and Materials Physics, Technical University Dresden, 01062 Dresden, Germany}
\affiliation{Leibniz Institute of Solid State and Materials Research (IFW Dresden), Helmholtzstr. 20, 01069 Dresden, Germany}

\author{H.~Reichlov\'a}
\affiliation{Institute of Solid State and Materials Physics, Technical University Dresden, 01062 Dresden, Germany}

\author{J.~\v{Z}elezn\'y}
\affiliation{Institute of Physics, Academy of Sciences of the Czech Republic, Cukrovarnick\'a 10, 162 00 Praha 6, Czech Republic}

\author{D.~Kriegner}
\email[]{kriegner@fzu.cz}
\affiliation{Institute of Solid State and Materials Physics, Technical University Dresden, 01062 Dresden, Germany}
\affiliation{Institute of Physics, Academy of Sciences of the Czech Republic, Cukrovarnick\'a 10, 162 00 Praha 6, Czech Republic}

\date{\today}

\begin{abstract}
The anomalous Hall effect, commonly observed in metallic magnets, has been established to originate from the time-reversal symmetry breaking by an internal macroscopic magnetization in ferromagnets or by a non-collinear magnetic order. Here we observe a spontaneous anomalous Hall signal in the absence of an external magnetic field in an epitaxial film of MnTe, which is a semiconductor with a collinear antiparallel magnetic ordering of Mn moments and a vanishing net magnetization. The anomalous Hall effect arises from an unconventional phase with strong time-reversal symmetry breaking and alternating spin polarization in real-space crystal structure and momentum-space electronic structure. The anisotropic crystal environment of magnetic Mn atoms due to the non-magnetic Te atoms is essential for establishing the unconventional phase and generating the anomalous Hall effect. 
\end{abstract}

\maketitle

The ordinary and anomalous Hall effects are prominent phenomena in condensed matter physics and spintronics that refer to a non-dissipative current, ${\bf j}_{\rm H}={\bf h}\times{\bf E}$, generated in a transverse direction to the applied electric field ${\bf E}$ \cite{Nagaosa2010,Smejkal2022}. The corresponding antisymmetric transverse components of the conductivity tensor, $\sigma^{\rm a}_{ij}=-\sigma^{\rm a}_{ji}$,  form a  pseudo-vector ${\bf h} = (-\sigma^{\rm a}_{yz}, \sigma^{\rm a}_{xz}, -\sigma^{\rm a}_{xy})$ whose components flip sign under the time-reversal transformation ($\cal{T}$) \cite{Nagaosa2010,Smejkal2022}. In the ordinary Hall effect, {\bf h} is typically proportional to an applied external magnetic field that breaks the $\cal{T}$-symmetry and couples to the electronic degrees of freedom via the Lorentz force. In contrast, the anomalous Hall effect (AHE) reflects the $\cal{T}$-symmetry breaking in the crystal and electronic structure associated with an internal magnetic order in a material \cite{Nagaosa2010,Smejkal2022}. The AHE can, therefore, generate a spontaneous Hall signal even in the absence of an external magnetic field. It is established that the AHE can arise from a macroscopic magnetization in ferromagnets, or a non-collinear order of spins on certain lattices of magnetic atoms \cite{Nagaosa2010,Smejkal2022}. 

Following a prediction for RuO$_2$ \cite{Smejkal2020}, the research of the AHE has recently turned to materials belonging to the abundant family of crystals with compensated collinear antiparallel ordering of spins \cite{Smejkal2020,Samanta2020,Naka2020,Mazin2021,Smejkal2021b}, for which the AHE and other $\cal{T}$-symmetry breaking spintronic phenomena have been traditionally considered to be excluded due to the vanishing net magnetization \cite{Smejkal2022}. The surprising emergence of the $\cal{T}$-symmetry breaking responses in these compensated magnetic crystals has been recently ascribed to an unconventional magnetic phase \cite{Smejkal2021b,Smejkal2022b}. It is 
characterized by an alternating spin polarization in both real-space crystal structure and momentum-space electronic structure, which has suggested to term this unconventional phase altermagnetism \cite{Smejkal2021b,Smejkal2022b}. Crystals hosting altermagnetism have the opposite-spin sublattices connected by crystal rotation transformations (proper or improper and symmorphic or non-symmorphic). 
In contrast, conventional antiferromagnets are characterized by the translation or inversion symmetry connecting the opposite-spin sublattices \cite{Nunez2006, Smejkal2017}. A systematic symmetry based classification and description of this unconventional magnetic phase has been developed using a non-relativistic spin-group formalism \cite{Smejkal2021b,Smejkal2022b}. 

RuO$_2$ is a prominent member of the altermagnetic family because of its metallic conduction combined with magnetic ordering above room temperature \cite{Berlijn2017,Zhu2019}, and because of the presence of the AHE \cite{Smejkal2020,Feng2020a} and of non-relativistic spin splitting, charge-spin conversion and giant magnetoresistive phenomena \cite{Smejkal2020,Ahn2019,Hayami2019,Feng2020a,Gonzalez-Hernandez2021,Smejkal2022c,Shao2021a,Bose2022,Bai2022,Karube2021}. Moreover, RuO$_2$ is a representative of the specific sub-class of altermagnets in which an anisotropic arrangement of non-magnetic atoms in the crystal (O-atoms in RuO$_2$) breaks the inversion and translation symmetry, while preserving the rotation symmetry, connecting the opposite-spin sublattices. Because of the essential role of non-magnetic atoms, the AHE in materials like RuO$_2$ has been referred to as originating from a crystal $\cal{T}$-symmetry breaking mechanism \cite{Smejkal2020,Samanta2020,Mazin2021}. 

Transport measurements in RuO$_2$ have confirmed the presence of the AHE signals \cite{Feng2020a}. However, the experiments relied on an externally applied magnetic field that served as a tool for reorienting the magnetic order vector from the [001] easy axis. Incidentally, this magnetic easy axis is a singular direction of the magnetic order vector in RuO$_2$ for which the AHE vanishes by symmetry \cite{Smejkal2020,Feng2020a}. 

In this Letter we report the experimental observation of the AHE in thin-film $\alpha$-MnTe with a room temperature compensated collinear magnetic order \cite{Kriegner2016}. Despite the extensive exploration of MnTe in the past \cite{greenwald1953, komatsubara1963, kunitomi1964,allen1977,przezdziecka2005,magnin2012,Kriegner2016,kriegner2017,bossini2021}, the AHE has not been theoretically or experimentally identified in this material prior to our work. We highlight that the MnTe structure shares with RuO$_2$ the essential role of non-magnetic atoms (Te atoms in case of MnTe) in the formation of the altermagnetic phase, and the corresponding $\cal{T}$-symmetry breaking phenomena, including the AHE. In contrast to RuO$_2$, our measurements in MnTe show hysteresis and spontaneous AHE signals at zero magnetic field. By symmetry analysis and density functional theory (DFT) calculations we show that the spontaneous nature of the AHE results from the favorable magnetic easy axes in our MnTe thin-film.

The other major distinction of MnTe from metallic RuO$_2$ and common ferromagnets or non-collinear magnets with metallic conduction is that stoichiometric MnTe is an intrinsic room-temperature magnetic semiconductor \cite{komatsubara1963,zanmarchi1967,allen1977,ferrer-roca2000,Kriegner2016}. 
The prospect of integrating $\cal{T}$-symmetry breaking spintronic phenomena with field-effect transistor functionalities in one material has driven the research of ferromagnetic semiconductors for decades, in particular of the III-V compounds that become ferromagnetic when heavily doped with Mn \cite{Ohno1998,Dietl2014,Jungwirth2014}. However, the required heavy charge dopings compromise the semiconducting nature of the host compounds, while the ferromagnetic transition temperatures have remained well below room temperature, i.e. far below the Curie temperatures of common transition-metal ferromagnets. Our observation of the AHE in MnTe demonstrates that altermagnetism can remove the roadblock associated with the notorious incompatibility of robust high-temperature ferromagnetism with semiconducting (insulating) band structures.

A schematic crystal of $\alpha$-MnTe with NiAs structure (crystallographic space group $P6_3/mmc$ \#194 \cite{villars}) is shown in Fig.~\ref{fig:1}a. The magnetic moments on Mn have a parallel alignment within the $c$-planes and an antiparallel alignment between the planes which allows to avoid frustration in the magnetic order. The crystallographic translation or inversion transformations do not connect the opposite spin sublattices because of the non-magnetic Te atoms that occupy non-centrosymmetric positions and form octahedra around the Mn atoms. This symmetry-lowering arrangement of non-magnetic atoms around the magnetic atoms is reminiscent of RuO$_{2}$ \cite{Smejkal2020}. The specific realization is, however, different. In RuO$_2$, the O octahedra belonging to the opposite-spin sublattices share an edge and the two sublattices are connected by a crystallographic non-symmorphic four-fold rotation $C_4$ combined with half unit-cell translation. In MnTe, the two opposite-spin sublattices are connected by a non-symmorphic six-fold rotation $C_{6}$ combined with half-unit cell translation along the $c$-axis $\mathbf{t}_\text{\sfrac{1}{2}}$ (screw-axis), as highlighted by the blue and red shaded face sharing octahedra in Fig.~\ref{fig:1}a \cite{Smejkal2022, Note1}.

\begin{figure}[tb]
\includegraphics[width=0.7\linewidth]{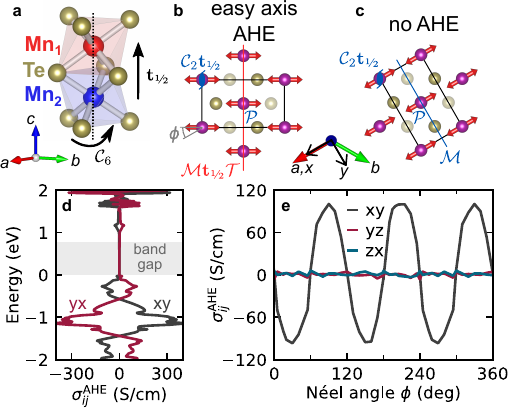}%
\caption{\label{fig:1} Theoretical calculation of spontaneous anomalous Hall signal in collinear MnTe: (a) Atomic configuration of Mn (blue/red) and Te (gold) with hexagonal NiAs structure. The two magnetic sublattices are connected by a six-fold screw axis along $[0001]$. (b,c) Magnetic moment configurations of the hexagonal $c$-planes with magnetic moments (red arrows) oriented along $\left[1\bar100\right]$ and $\left[2\bar1\bar10\right]$, respectively. The magnetic unit cell shape (black line) and crystal symmetry operations corresponding to the generators of the magnetic point groups are indicated in the panels. Te atoms at different heights are indicated by different color saturation. (d) DFT Anomalous Hall conductivity vs.\ Fermi energy calculated for the moment configuration of panel (b) and Hall current along $x$-direction. (e) Transverse conductivity components for the Fermi energy 0.25~eV below the valence band maximum as a function of the N\'eel vector orientation. The angle $\phi$ is defined with respect to the $a$-axis. 
}
\end{figure}

The strong non-relativistic $\cal{T}$-symmetry breaking and alternating spin-splitting in the band structure of MnTe, accompanied by zero non-relativistic net magnetization, is described by a spin Laue group $^2{6/}{^2}m^2m^1m$ of symmetry transformations in decoupled real and spin space \cite{Smejkal2021b}. The spin-group symmetries imply the presence of four spin-degenerate nodal planes crossing the $\boldsymbol\Gamma$-point of the momentum-dependent electronic structure ($g$-wave symmetry), corresponding to four mirror-symmetry planes combined with spin-space rotation of the compensated collinear spin arrangement on the real-space MnTe crystal \cite{Smejkal2021b}. 

Besides the $\cal{T}$-symmetry breaking, additional symmetry-lowering requirements have to be fulfilled by the magnetic state of the crystal to allow for an Hall pseudo-vector {\bf h} \cite{Smejkal2022}. These requirements are described within the formalism of relativistic magnetic-group symmetries in coupled real and spin space because, in general, the AHE in collinear (coplanar) magnets is observed in the presence of spin-orbit coupling \cite{Smejkal2022}. Whether or not the AHE is allowed by the relativistic magnetic symmetries then depends on the orientation of the magnetic-order (N\'eel) vector with respect to the crystal axes. 

Figures~\ref{fig:1}b,c show two representative orientations of the N\'eel vector within the $c$-plane of MnTe. The first orientation is along the $\left[1\bar100\right]$ magnetic easy axis in our thin-film \cite{Note1}, which is rotated by $\phi=30^{\circ}$ from the crystallographic $a$-axis \cite{Note2, kriegner2017}.
In this case, the generators of the relativistic magnetic point-group $m'm'm$ are the inversion $\cal{P}$, a two-fold rotation $C_{2}$ around the $c$-axis, and a mirror perpendicular to the N\'eel vector combined with time-reversal $\cal{M}\cal{T}$ (indicated in Fig.~\ref{fig:1}b). This is compatible with the presence of the AHE pseudo-vector along the $c$-axis \cite{Note1,Smejkal2020,Smejkal2022} and therefore enables the detection of a hysteretic AHE signal.
In contrast, for the N\'eel vector along the $a$-axis ($\left[2\bar1\bar10\right]$),
the generators of the relativistic magnetic point group $mmm$ are inversion $\cal{P}$, the two-fold rotation $C_{2}$ around $c$, and a mirror $\cal{M}$ perpendicular to the N\'eel vector, which excludes the AHE by symmetry \cite{Note1}. 

To calculate the intrinsic Berry-curvature anomalous Hall conductivity (AHC) \cite{Nagaosa2010,Smejkal2022}, we use the maximally localized Wannier functions based on the DFT framework \cite{Note1}. The obtained AHC for various N\'eel vector orientations within the $c$-plane corroborate our symmetry analysis.
For the N\'eel vector aligned with the easy axis and current along the $a$-axis, the dependence of the AHC on the Fermi energy is shown in Fig.~\ref{fig:1}d. AHC values above 300~S/cm are found for states deep in the valance band, while the magnitude diminishes when the Fermi level approaches the valance band edge. Note that for the indices of the components of the AHE pseudo-vector ${\bf h} = (-\sigma^{\rm AHE}_{yz}, \sigma^{\rm AHE}_{xz}, -\sigma^{\rm AHE}_{xy})$, we use a notation in which $x$ corresponds to the crystal $a$-axis ($\left[2\bar1\bar10\right]$), $y$ to the $\left[01\bar10\right]$ crystal axis, and $z$ to the $c$-axis ($\left[0001\right]$) of MnTe. 

Figure~\ref{fig:1}e shows the AHC components $\sigma_{ij}^{\rm AHE}$ calculated for a fixed Fermi energy as a function of the N\'eel vector angle $\phi$ in the $c$-plane measured from the $a$-axis. Here $i$ labels the Hall current direction and $j$ the applied electric-field direction. As expected from symmetry, $\sigma_{xy}^{\rm AHE}$ is the only non-zero component of the AHE pseudo-vector, i.e., {\bf h} is parallel to the $c$-axis and orthogonal to the in-plane N\'eel vector. 
This is strikingly distinct from the conventional AHE in ferromagnets where the Hall vector typically aligns with the magnetization order vector \cite{Nagaosa2010}. Similarly, it is distinct from the experimentally observed anomalous Hall effect in RuO$_2$, ascribed to the N\'eel vector component parallel to the Hall vector \cite{Feng2020a}.
The symmetry restrictions of MnTe also force $\sigma_{xy}^{\rm AHE}(\phi)$ to zero every 60 degrees, when the N\'eel vector is aligned along the $a$-axis or equivalent axes in the $c$-plane. 

For the AHE measurements, we have prepared epitaxial thin films of $\alpha$-MnTe by molecular beam epitaxy on a single-crystal InP(111)A substrate at a substrate temperature of 380 $^\circ$C using Mn and Te beam flux sources. Reflection high energy electron diffraction and X-ray diffraction data indicate two-dimensional layer-by-layer growth, high crystallographic quality, and film thickness of 48~nm \cite{Note1}.
For the growth on InP(111)A, the epitaxial relationship is $(0001)\left[1\bar100\right]_\textrm{MnTe}\parallel(111)\left[11\bar2\right]_\textrm{InP}$, which allows us to perform the desired AHE transport measurements in the $c$-plane. Charge transport in our thin films is enabled by unintentional p-type doping commonly observed in MnTe \cite{wasscher1969,ferrer-roca2000}. Details of the growth, transmission electron microscopy, and neutron diffraction studies of our films are described by Kriegner and coworkers~\cite{Kriegner2016,kriegner2017}. 

Hall bar microdevices, shown in Fig.~\ref{fig:2}a, were fabricated using electron beam lithography, argon milling, and Cr/Au contacts produced by a lift-off process. 
The Hall bar is along the $y$-axis ($\left[01\bar10\right]$), and the magnetic field is applied in the $\left[01\bar10\right]-\left[0001\right]$ plane at an angle $\beta$ measured from the Hall-bar direction. In our semiconducting MnTe, the transverse resistivity plotted in Fig.~\ref{fig:2}b is dominated by the ordinary Hall contribution that is a linear function of the out-of-plane ($\left[0001\right]$) component of the applied magnetic field. The hole concentration and mobility inferred from the measurements in Fig.~\ref{fig:2} by assuming a single band picture are $4.9\times10^{18}$ cm$^{-3}$ and $29$ cm$^2$/Vs, consistent with the semiconducting character of MnTe and with a Fermi level within $\sim 20$~meV from the valance band edge.

\begin{figure}[tb]
\includegraphics[width=\linewidth]{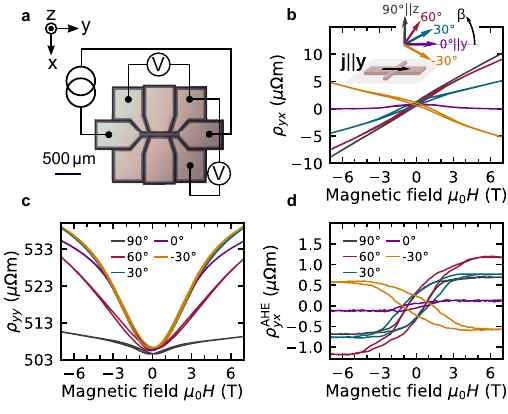}%
\caption{\label{fig:2} Magnetic field sweep measurements and AHE at $T = 150$ K. (a) Microscopy image of the Hall bar together with electrical schematics of our measurement. For all transport measurements a moderate current density $j\sim 8\cdot10^6$~A/m$^2$ was used.
(b,c) Transverse and longitudinal resistivities measured during magnetic field sweeps in a geometry indicated by the sketch in (b). (d) Inferred anomalous Hall resistivity given by, $\rho_{yx}^{\rm AHE}\equiv\rho_{yx}-\rho_{yx}^{\rm OHE}-\rho_{yx}^{\rm even}$ (see text).
}
\end{figure}

Besides the dominant ordinary Hall signal, the transverse resistivity in Fig.~\ref{fig:2}b contains an anomalous contribution, as seen from the hysteresis with different zero-field spontaneous Hall signals for the opposite field-sweep directions. In contrast, such a hysteretic behavior is absent in the longitudinal resistivity measurement, as shown  in Fig.~\ref{fig:2}c. The anomalous Hall signal $\rho_{yx}^{\rm AHE}$ is highlighted in Fig.~\ref{fig:2}d where we subtracted the ordinary Hall contribution ($\rho_{yx}^{\rm OHE}$) from the measured transverse resistivity, and where we also removed the even-in-field contribution defined as $\rho_{yx}^{\rm even}=[\langle\rho_{yx}(H)\rangle+\langle\rho_{yx}(-H)\rangle ]/2$. Here  $\langle\rho_{yx}(H)\rangle=[\rho_{yx}(H_\uparrow)+\rho_{yx}(H_\downarrow)]/2$, and $H_{\uparrow(\downarrow)}$ refers to the measurement at the negative-to-positive (positive-to-negative) field sweep \cite{Note1}.

The hysteresis and the opposite zero-field spontaneous Hall signals for opposite field-sweep directions are well pronounced in $\rho_{yx}^{\rm AHE}$ when the applied magnetic field contains an out-of-plane component. This is again explained by symmetry. MnTe can have a weak relativistic magnetization due to terms in the thermodynamic potential that couple the weak magnetization to the N\'eel vector \cite{Landau1984}. The allowed presence of the weak magnetization in zero magnetic field requires in our MnTe the same symmetry lowering as the spontaneous AHE because both effects are described by a pseudo-vector that is odd under $\cal{T}$ \cite{Smejkal2022}. For the N\'eel vector in the $c$-plane of MnTe, the anomalous Hall and the weak magnetization pseudo-vectors align with the $c$-axis. Except when the N\'eel vector is parallel to the $a$-axis or the other equivalent axes for which both effects vanish by symmetry. When an applied out-of-plane magnetic field reverses the weak magnetization, symmetry implies that the N\'eel vector is also forced via the coupling in the thermodynamic potential to reverse, which results in the flip of the sign of the AHE.

The presence of the in-plane component of the magnetic field in our measurements is favorable for maximizing the observed $\rho_{yx}^{\rm AHE}$ because of the repopulation of domains with the N\'eel vectors aligned with the three equivalent zero-field easy axes. The repopulation, i.e., the in-plane N\'eel vector reorientation occurs for in-plane fields above the spin-flop transition which is around 2~T \cite{Note1}. As a result, the measured AHE at saturation is maximized for $\beta = 60^\circ$, while $\beta = 90^\circ$ gives the lowest saturation field (see Fig.~\ref{fig:2}d).

Note that the decisive role of the out-of-plane field component for determining the sign of the spontaneous and saturated AHE signal is particularly visible for $\rho_{yx}^{\rm AHE}$ traces recorded at $\beta=\pm30^\circ$. While the in-plane component of the field is the same in both of these field-sweep traces, the out-of-plane components have the opposite sign and, correspondingly, the measured hysteretic $\rho_{yx}^{\rm AHE}$  also flips sign between the two traces. When removing the magnetic field, some domain randomization occurs due to magnetostriction \cite{Kriegner2016}. This causes a reduction of the spontaneous zero-field AHE signal as compared to the saturated value (Fig.~\ref{fig:2}d). 

We also point out that while the weak magnetization facilitates the reversal of the sign of $\rho_{yx}^{\rm AHE}$ by the out-of-plane component of the applied magnetic field, the AHE is not a consequence of the weak magnetization, as discussed in detail by L. \v{S}mejkal and coworkers~\cite{Smejkal2020,Smejkal2022}. The principle origin of the AHE is the ${\cal T}$-symmetry breaking due to the compensated antiparallel magnetic order on Mn atoms and the anisotropic crystal environment due to the Te atoms. The relativistic spin-orbit interaction then couples this strong ${\cal T}$-symmetry breaking to the charge Hall transport and, analogous but not as a consequence of the other, can also generate the weak magnetization \cite{Smejkal2020,Smejkal2022}. 

Measurements of the magnetization, shown in Fig.~\ref{fig:3}a, are consistent with the above origin of the spontaneous AHE in MnTe, and contrast with the conventional magnetization-induced AHE in ferromagnets. We performed the magnetization measurements using the superconducting quantum interference device (SQUID) technique. The raw SQUID signal is dominated by the diamagnetic contribution from the substrate. After subtracting a linear background, we observe that the remanent out-of-plane magnetization, albeit allowed by symmetry and estimated to be $\lesssim 2\times 10^{-4}$ $\mu_\mathrm{B}$/Mn from first principle calculations, is unobservable even within the high-resolution of the employed SQUID technique. This contrasts with the clearly observable spontaneous zero-field signals and hysteresis in the AHE measurements. Note that for in-plane magnetic fields, the kink found in the in-plane magnetization between 2 and 3~T (Fig.~\ref{fig:3}a) marks the spin-flop reorientation field of the N\'eel vector into an in-plane direction orthogonal to the applied field. This is visible for any inplane direction since three equivalent easy axes exist in the basal plane \cite{kriegner2017}. First principle calculations let us also exclude the weak magnetization as source for the found AHE since significantly larger magnetization would be needed to yield a sizable AHE. 

\begin{figure}[tb]
\includegraphics[width=\linewidth]{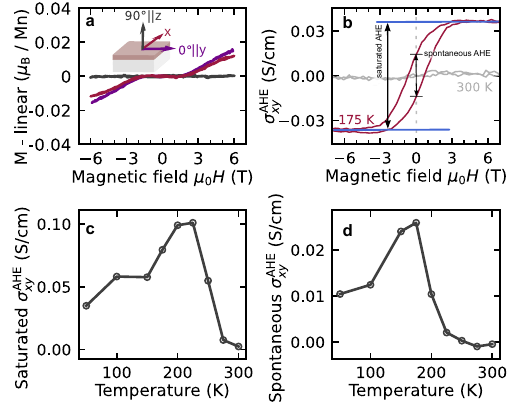}
\caption{\label{fig:3} (a) SQUID magnetization obtained for in-plane and out-of-plane magnetic field measured at 10~K on a 2000~nm thick film. To highlight the absence of remanent magnetization a linear slope obtained in vicinity of zero magnetic field was subtracted from each data set. Note that the slope contains the susceptibility of MnTe as well as the diamagnetic substrate. (b) AHE conductivity $\sigma_{xy}^{\rm AHE}$ obtained from the measured $\rho_{yx}^{\rm AHE}$ at 175 and 300~K, and $\beta=30^\circ$. (c,d) Temperature dependence of the spontaneous zero-field and saturated AHE conductivity.
}
\end{figure}

In Fig.~\ref{fig:3}b we show the conversion of $\rho_{yx}^{\rm AHE}$ at 175 and 300~K, and $\beta=30^\circ$ into the anomalous Hall conductivity $\sigma_{xy}^{\rm AHE}$. The temperature-dependent saturated and spontaneous AHE conductivity signals shown in Figs.~\ref{fig:3}c,d were extracted from $\sigma_{xy}^{\rm AHE}$ as indicated in Fig.~\ref{fig:3}b. As expected for the AHE, they vanish when approaching the MnTe N\'eel temperature of 310~K. The complex temperature dependence below the N\'eel temperature can be governed by a combination of the variation of the longitudinal resistivity \cite{Note1, magnin2012}, the achievable degree of polarization of the system, as well as a potential variation of the Fermi level. Nevertheless the measured magnitude of $\sigma_{xy}^{\rm AHE}$ is within the scale predicted by our DFT calculations for the Fermi level on the order of $\sim 20$~meV below the valence-band edge, consistent with the hole density inferred from the ordinary Hall signal and using finite band broadening \cite{Note1}.

Finally, we note that hysteretic and spontaneous AHE signals are also found for the electrical current applied along the in-plane $a$-axis ($\left[2\bar1\bar10\right]$). This is consistent with the out-of-plane $c$-axis orientation of the AHE pseudo-vector that allows for the AHE detection with any in-plane applied current.

In conclusion, we have detected a spontaneous AHE in semiconducting MnTe with collinear compensated magnetic order. In magnetic field sweeps, we identify open and saturated hysteresis loops of the AHE signal, consistent with the $\cal{T}$-symmetry breaking by the antiparallel order of spins and anisotropic crystal environment in MnTe, and in contrast to the conventional AHE in ferromagnets generated by a macroscopic net magnetization. Our theoretical symmetry analysis and DFT calculations corroborate the experimental results. They confirm the favorable crystallographic orientation of the easy axis (axes) in MnTe films for generating the spontaneous AHE signal. DFT calculations of the intrinsic Berry-curvature contribution to the AHE conductivity are consistent with the measured scale of the AHE signals. Our work opens the prospect of exploring and exploiting $\cal{T}$-symmetry breaking phenomena arising from the unconventional altermagnetic phase in semiconductors and insulators. 

\begin{acknowledgments}
We acknowledge M. Rahn for his support with the crystal direction determination.
This work was supported in part by the Ministry of Education of the Czech Republic Grants LM2018110 and LNSM-LNSpin, by the Grant Agency of the Czech Republic Grant No. 22-22000M, by the Deutsche Forschungsgemeinschaft (DFG, German Research Foundation) - TRR 173/2 SPIN+X - 268565370 (project A03), by the French national research agency (ANR) (Project MATHEEIAS, Grant No. 445976410), by the Austrian Science Funds - Project P30960-N27, and by the EU FET Open RIA Grant No. 766566. Part of the experiments were performed at MGML (mgml.eu), which is supported within the program of Czech Research Infrastructures (Project No. LM2018096). D.K. acknowledges the support by the Lumina fellowship LQ100102201 of the Czech Academy of Sciences. We acknowledge the computing time granted on the supercomputer Mogon at Johannes Gutenberg University Mainz (hpc.uni-mainz.de). 
\end{acknowledgments}

\footnotetext[1]{See Supplemental Material at [URL] for additional experimental data and figures.}
\footnotetext[2]{Note that we use Bravais indices $hkil$, with $i = -h-k$ for the hexagonal MnTe, while Miller indices $hkl$ are used for the cubic substrate}

\appendix
\renewcommand\thefigure{S\arabic{figure}} 
\setcounter{figure}{0}

\pagebreak

\pagebreak
\title{Spontaneous anomalous Hall effect arising from an unconventional compensated magnetic phase in a semiconductor\\
---\\
Supplementary information}

\maketitle

\section{Magnetic unit cells of MnTe}

\begin{figure}[htb]
\includegraphics[width=\linewidth]{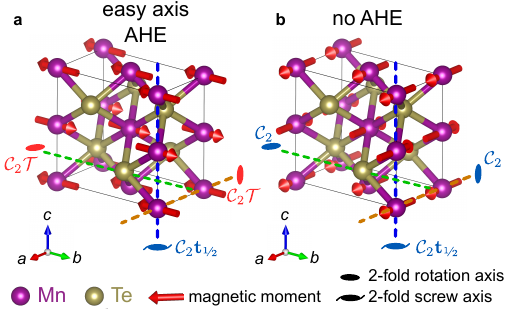}
\caption{\label{fig:unitcell} Magnetic unit cells of MnTe for two magnetic moment configurations. In panel (a) the magnetic moments are rotated from the hexagonal $a$-axis by 30 degree. (b) Magnetic moments oriented along the hexagonal $a$-axis of the crystalline unit cell. The hexagonal crystalline symmetry is reduced to an orthorhombic magnetic unit cell in both cases which is indicated by black lines. A subset of the symmetry operations consisting of two-fold rotation and screw axis are indicated. Vectors $a$, $b$, $c$ indicate the magnetic unit cell vectors for the two configurations. Note that $a$/$c$ of the magnetic orthorhombic and crystalline hexagonal cell coincide.
}
\end{figure}

Here we briefly summarize the main symmetry properties for the two magnetic moment configurations shown in Fig.~1b,c of the main text. Figure~\ref{fig:unitcell} shows these two collinear magnetic order configurations in 3D. The two magnetic unit cells have the moments oriented along $\left[1\bar100\right]$ (30 degree rotated from $a$) and $\left[2\bar1\bar10\right]$ (along $a$) of the paramagnetic hexagonal unit cell. The magnetic cells for both orientations are described by an orthorhombic unit cell, which is a subgroup of the hexagonal paramagnetic cell. The orthorhombic cell has its $a$ and $c$ axis along with the corresponding axis of the crystalline hexagonal cell. Figure~\ref{fig:unitcell}a with the magnetic moments along the experimentally reported easy axis has magnetic space group symmetry $Cm'c'm$ (\#63.462) while the orthogonal arangement of the moments (Fig.~\ref{fig:unitcell}b) corresponds to $Cmcm$ (\#63.457). The difference of the symmetry properties of the unit cells are examplified in Fig.~\ref{fig:unitcell} by marked rotation axes. In the case of the easy axis orientation ($Cm'c'm$) a Hall pseudovector can exist along the $c$ axis since the orthogonal rotation axis in combination with the time reversal operation ($\mathcal{T}$) leave it unchanged. Note that also all other not indicated symmetry operations do not alter a pseudovector along $c$. Other components of a pseudovector need to vanish because of the two fold screw axis along $c$. In contrast no pseudovector components, and therefore no AHE, can exist for the $Cmcm$ symmetry since the three orthogonal rotation axis would always transform any vector into a different state and thereby forcing it to vanish. The difference of the two magnetic symmetries manifests itself also in the fact that only the $Cm'c'm$ symmetry is compatible with ferromagnetism. Note that in addition to the marked rotation axes, the structures contain mirror plane(s) and centers of inversion which do not change the conclusion presented above.

\FloatBarrier
\section{Spin split band structure of MnTe}

First-principles calculations based on density-functional theory (DFT) were used to study the electronic structure and transport properties of MnTe. We have used the VASP (Vienna $ab$-$initio$ simulation package) code \cite{vasp1996} and GGA+U (we use $U=$3.03~eV to reconstruct well previous theory and experimental results\cite{kriegner2017a,yin2019,Mu2019}). The electron wave function was expanded in plane waves up to cut-off energy of 500 eV and a 12$\times$12$\times$16 $k$-mesh has been used for the Brillouin-zone integration. By using maximally localized Wannier functions \cite{wannier90,Pizzi2020}, we have found an effective tight-binding Hamiltonian to calculate the anomalous Hall conductivity tensor. Integration of the Kubo formula was carried out within a $400^3$ $k$-grid with the Wannier Tools code \cite{wanniertools}. 

\begin{figure}[htb]
\includegraphics[width=\linewidth]{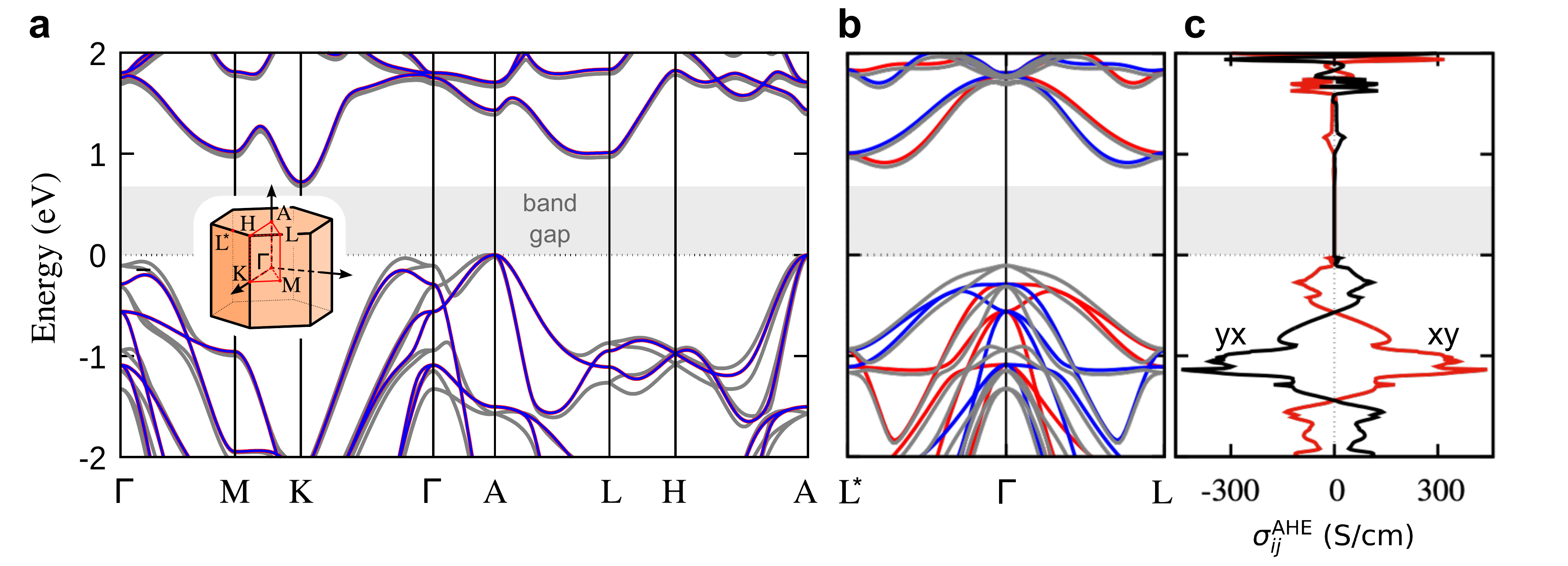}%
\caption{\label{fig:bands} Band structure of MnTe. (a) Calculated bands near the Fermi energy. Along high symmetry directions bands for the two spin channels (shown in blue and red) overlap. The used high symmetry points are labelled in the inset and the band gap of the semiconducting material is marked. (b) Between the $\Gamma$ and $L$ direction bands for different spin channels show significant splitting. Gray lines in (a,b) shows the band structure calculation including the effect of spin orbit coupling. (c) The energy dependent anomalous Hall conductivity calculated from the shown band structure.
}
\end{figure}

Our VASP band structure calculations of MnTe are shown in Fig.~\ref{fig:bands}. Along the commonly shown hexagonal high symmetry directions (left panel) in agreement with previously published band structure calculations \cite{podgorny1983, wei1987, yin2019} we find no spin splitting. However, when looking between the $\Gamma$ and $L$ points a splitting of spin polarized bands is observed which amounts to several 100 meV. It's this splitting of the spin polarized bands which causes the discussed anomalous Hall effect.

\begin{figure}[htb]
\includegraphics[width=0.7\linewidth]{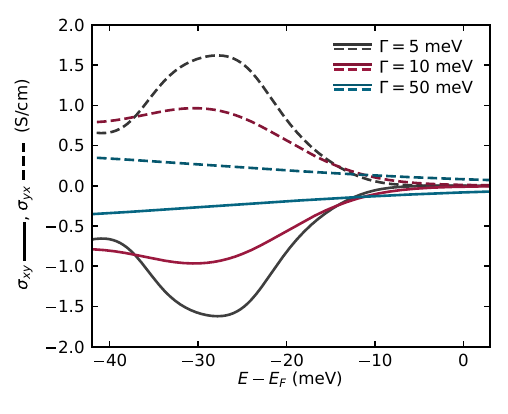}%
\caption{\label{fig:ahe_gamma} The energy dependent anomalous Hall conductivity obtained for various band broadening values in the vicinity of the Fermi level.
}
\end{figure}

While the calculated anomalous Hall conductivity is sizable deep in the valence band it approaches zero at the Fermi level. Based on the experimental measurement of the hole concentration, we estimate the Fermi level to be shifted by $\sim20$ meV from the top of the valence band. This estimate carries some uncertainty, however, since the experimental estimate is based on a single band picture. For this position of the Fermi level we find anomalous Hall conductivity $\sim 1$ S/cm. Taking the uncertainty of the Fermi level position into account, the estimated anomalous Hall conductivity is $\sim 0.1 - 10 $ S/cm. This is a calculation of the zero-temperature intrinsic anomalous Hall conductivity, which corresponds to a perfect MnTe crystal. In a real system, there are be impurities and imperfections, which will change the AHE value. This can be simulated by including a band broadening $\Gamma$ in the calculation. Here we use the Kubo formula derived in Ref. \cite{freimuth2014} and the code \cite{linear_response_code}. The result for several values of $\Gamma$ and for a range of Fermi level values close to the top of the valence band is shown Fig.~\ref{fig:ahe_gamma}. It is seen that even a modest $\Gamma$ results in a decrease of the AHE in the whole energy range, bringing the results quite close to the experimentally observed values.

\FloatBarrier
\section{Thin film growth and characterization}

Epitaxial MnTe layers were grown in a Varian GEN II molecular beam epitaxy system. The films were deposited onto InP(111)A substrates after oxide desorption at a growth temperature of 380 °C. The MnTe growth rate was around 1 \AA/sec and an excess Te flux of 1 monolayers/sec was used, with the Mn and Te flux rates calibrated by the quartz microbalance method. The growth was monitored in situ by reflection high-energy electron diffraction (RHEED) and the perfectly streaked diffraction patterns indicate a 2D growth of the layer. 

\begin{figure}[htb]
\includegraphics[width=\linewidth]{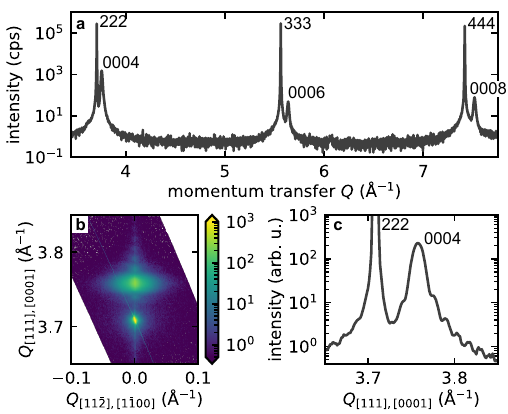}%
\caption{\label{fig:xrd}  X-ray diffraction data of MnTe thin films on InP(111)A substrate. (a) Radial scan evidencing the epitaxial growth of the thin films. Only $HHH$ and $000L$ diffraction maxima of InP and MnTe are observed. (b) Reciprocal space map near the 222 InP and 0004 MnTe diffraction maximum. Note that the intensity scale is saturated at the substrate peak. Laue thickness fringes along the out of plane direction are visible. Those are even more evident in a vertical line cut taken in this reciprocal space map, which is shown in panel (c).
}
\end{figure}

X-ray diffraction data of MnTe thin films on InP(111)A substrate are shown in Fig.~\ref{fig:xrd}. A symmetric radial scan shown in Fig.~\ref{fig:xrd}a probing the lattice planes parallel to the sample surface finds only $HHH$ and $000L$ diffraction maxima of InP and MnTe, respectively. The two-dimensional diffraction intensity around the InP 222 and MnTe 0004 Bragg diffraction peaks is mapped out in Fig.~\ref{fig:xrd}b. A low mosaicity of the thin film leads to a narrow peak. The high crystalline quality is evident from the visibility of Laue thickness fringes along the out of plane $[0001]$ direction, which confirm the presence of a coherent lattice over the full film thickness. Fig.~\ref{fig:xrd}c shows a line cut through the diffraction map which highlights these thickness fringes and allows us to determine the film thickness which is found to be 48~nm.

\begin{figure}[tb]
\includegraphics[width=\linewidth]{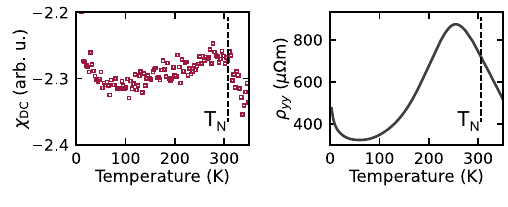}%
\caption{\label{fig:rho_chi} Temperature dependent resistivity and DC susceptibility. The bulk N\'eel temperature $T_N$ is indicated at 310~K. The susceptibility measurements were performed on XX~nm thick MnTe films in a field of 0.05~T. The upturn in susceptibility at low temperatures can be attributed to paramagnetic oxygen inside the magnetometer or contribution from paramagnetic impurities in the substrate. 
}
\end{figure}

Similar to previous studies we find the longitudinal resistivity with the characteristic temperature variation typical for MnTe (Fig.~\ref{fig:rho_chi}a). Although MnTe is a semiconductor with $\sim1.4$~eV band gap \cite{zanmarchi1967, allen1977, ferrer-roca2000} the resistivity is decreasing below 250~K and only raises again below 50~K. This was previously attributed to spin dependent scattering processes \cite{magnin2012}. In our case the decrease of $\rho_{yy}$ at temperatures above 250~K is governed by the thermal activation of carriers in the semiconducting InP substrate and can not be solely attributed to the MnTe thin film. Magnetotransport studies therefore focus on lower temperatures where the substrate can be considered isolating.
The compensated magnetic nature of these thin films can be seen from the low susceptibility variation shown in Fig.~\ref{fig:rho_chi}b, which only shows a weak anomaly near the expected magnetic transition temperature and is otherwise dominated by the diamagnetic substrate. 

\FloatBarrier
\section{Symmetry analysis of the conductivity/resistivity tensor}
\label{sec:tensorsym}

The calculations presented in the main text's figure 1 only considered the AHE, but in experiments we expect to find also other contributions to the conductivity/resistivity tensor upon a N\'eel vector rotation. Symmetry considerations outlined below lead us to expect a series of anisotropic magnetoresistance (AMR) contributions. In particular when considering terms up to the sixth-order the conductivity $\sigma_{yy}$ (in the coordinate system of Fig.~1 of the main text) is expected to contain terms proportional to $\cos n\phi$, with $n=2,4,6$. The transversal conductivity $\sigma_{yx}$ can contain terms proportional to $\sin n\phi$, with $n=2,3,4$.

To analyze the symmetry of the conductivity tensor, we expand it in the N\'eel vector $\mathbf{L}$:

\begin{align}
    \sigma_{ij}(\mathbf{L}) = \sigma_{ij}^{(0)} + \sigma_{ijk}^{(1)} L_k + \sigma_{ijkl}^{(2)} L_l L_l + \dots
\end{align}

We determine the symmetry of each term in this expansion, using the code Symmetr \cite{symcode}. In the experiments, $\mathbf{L}$ is expected to be always in the $ab$ plane and thus can be expressed as $\mathbf{L}/L = (\cos(\phi),\sin(\phi),0)$, where $\phi$ denotes the angle between the electric field and $\mathbf{L}$. Substituting this for $\mathbf{L}$ we have an expansion in powers of $\cos(\phi)$ and $\sin(\phi)$. This can be converted to expansion in $\cos(n\phi)$ and $\sin(n\phi)$, where $n$ are integers. Note that in general terms of $n$-th order in the powers of $\cos(\phi)$ and $\sin(\phi)$ expansion correspond to $n$-th and \emph{lower} order terms in the $\cos(n\phi)$ and $\sin(n\phi)$ expansion.  Considering current along the $y$-direction, terms up to sixth order, and separating the components parallel and perpendicular to current, we obtain:

\begin{align}
    \sigma_{yy} &= \sigma^{(0)} + \sigma^{(2)} \cos(2\phi) + \sigma^{(4)} \cos(4\phi) + \sigma^{(6)} \cos(6\phi), \\
    \sigma_{yx} &= \sigma^{(2)} \sin(2\phi) + \sigma^{(3)} \sin(3\phi)- \sigma^{(4)} \sin(4\phi),
\end{align}

where $\sigma^{(n)}$ corresponds to the amplitude of the respective conductivity contribution.
Qualitatively similar results are obtained for current in the $x$-direction and the results apply analogously to the resistivity tensor.

\FloatBarrier
\section{Even/odd separation in magnetic field sweeps}
\label{sec:even_odd}

In the main text, we show measurements of $\rho_{xy}(H)$ for the magnetic field swept from negative to positive ($H\uparrow$) and positive to negative ($H\downarrow$). The average of both sweep directions $\left<\rho_{xy}(H)\right> = \left(\rho_{xy}(H\uparrow) + \rho_{xy}(H\downarrow)\right)/2$ is used to define the component symmetric, i.e. even, in magnetic field
\begin{equation}
    \rho_{xy}^{\rm even}(H) = \frac{\left<\rho_{xy}(H)\right> + \left<\rho_{xy}(-H)\right>}{2}.
\end{equation}
The odd contribution for up and down sweeps is then obtained by subtraction 
\begin{equation}
\label{eq:rhoodd}
    \rho_{xy}^{\rm odd}(H) = \left\{
    \begin{array}{r}
        \rho_{xy}(H\uparrow) - \rho_{xy}^{\rm even}(H) \quad\text{for } H\uparrow \\
        \rho_{xy}(H\downarrow) - \rho_{xy}^{\rm even}(H) \quad\text{for } H\downarrow
    \end{array} \right. .
\end{equation}

In the odd signal obtained by Eq.~\ref{eq:rhoodd} also the odd in magnetic field ordinary Hall effect is contained. In order to highlight the hysteretic nature of the odd contribution a linear slope assigned to the ordinary Hall effect is typically subtracted.

\begin{figure}[tb]
\includegraphics[width=\linewidth]{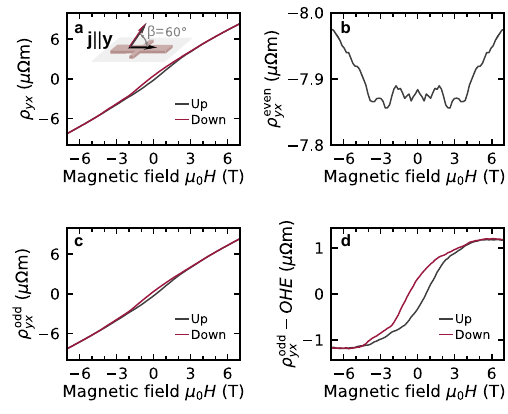}
\caption{\label{fig:even_odd} Transversal magnetoresistance for a magnetic field sweep recorded at $T= 150$~K with a magnetic field angle $\beta=60^\circ$ with respect to the current direction. (a) As measured transversal resistivity data. (b) Even in magnetic field signal $\rho_{yx}^{\rm even}$ obtained by symmetrizing the average of up and down field sweep. (c) Odd contribution $\rho_{xy}^{\rm odd}$ obtained from the raw data by subtraction of the even contribution. (d) Odd contribution $\rho_{yx}^{\rm odd}$ minus a linear contribution attributed to the ordinary Hall effect.
}
\end{figure}

Figure~\ref{fig:even_odd} shows the various steps in the separation of even and odd signals and how the hysteretic contribution is isolated. 

\FloatBarrier
\section{Additional magnetotransport data}
\label{sec:more_exp}

\begin{figure}[htb]
\includegraphics[width=\linewidth]{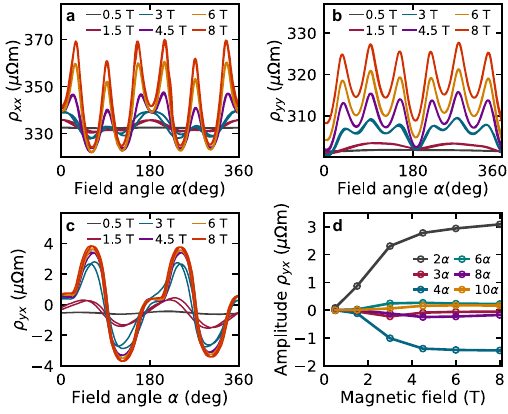}%
\caption{\label{fig:mrrot} Angle dependent magnetotransport data recorded at $T=50$~K. (a,b) Longitudinal resistivity variation for inplane magnetic field for current along the $x\parallel\left[2\bar1\bar10\right]$ and $y\parallel\left[01\bar10\right]$ directions. (c) Transversal resistivity variation for inplane magnetic field and current along $y$. The magnetic field dependence of the leading harmonic contributions to data in (c) is given in panel (d). The field angle $\alpha$ is defined with respect to the crystalline $a$-axis (c.f. Fig.~1 of the main text).
}
\end{figure}

Complementary to the magnetic field sweep measurements shown in the main text we show here angle dependent longitudinal and transveral resistivities for inplane magnetic fields.
We rotate the magnetic field with various strength in the field plane and obtain the traces shown in Fig.~\ref{fig:mrrot}. Various harmonic contributions are observed which tend to saturate at the highest fields. Analysis of the data shows that the strongest contributions to the longitudinal resistivity (Fig.~\ref{fig:mrrot}a,b) depend as predicted on $\cos(n\phi)$ with $n=2,4,6$. These terms are even in the magnetic field and are attributed to the anisotropic magnetoresistance (AMR). At high fields, the six-fold contribution is strongest. It is noteworthy that the six-fold contributions in $\rho_{xx}$ and $\rho_{yy}$ follow the same pattern with minima observed at the same angles (0, 60, 120$^\circ$ ,\dots) whereas the two-fold contribution changes its sign and has a maximum at 0$^\circ$ in $\rho_{xx}$ while a minimum is observed at the same angle in $\rho_{yy}$. This indicates that the two-fold contribution is depending on the angle between N\'eel vector and current (non-crystalline AMR), whereas the six-fold contribution depends on the angle between N\'eel vector and crystal axis (crystalline AMR).

The angle dependence of the transversal resistivity is shown in Fig.~\ref{fig:mrrot}c and is dominated by the even contributions with two-fold and four-fold symmetry. In contrast to the longitudinal resistivities, the six-fold contribution is absent in this geometry. This is in agreement with the AMR contributions derived in Sec.~\ref{sec:tensorsym}. It also is in line with results in cubic/tetragonal systems, where crystalline AMR terms did not contribute to the transversal resistivities \cite{ranieri2008}. The amplitudes of the main contributions (two- and four-fold) do, however, agree with what was found in the corresponding longitudinal measurement.

\begin{figure}[htb]
\includegraphics[width=\linewidth]{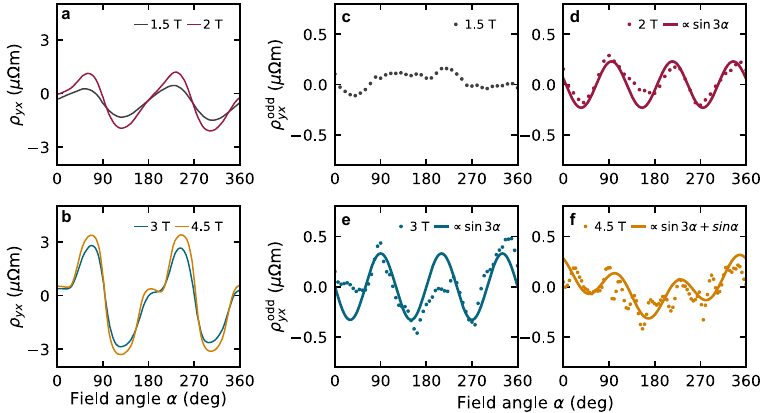}%
\caption{\label{fig:mr3fold} Angle dependent transversal magnetotransport data recorded at $T=50$~K for inplane magnetic field and current along $y$ ((excerpt from Fig.~\ref{fig:mrrot}c). (a, b) Transversal resistivity variation at field strength of 1.5 \& 2 T and 3 \& 4.5 T. (c-f) Odd in magnetic field contributions obtained after subtraction of fitted $sin(n\alpha)$ with even $n$ ($n=2,4,6,8,10$). A guide to the eye proportional to $\sin 3\alpha$ is shown in (d, e). The appearance of an additional contribution proportional to $\sin\alpha$ in data with higher fields is due to wobbling of the axis of the mechanical rotator which causes an unintentional out of plane field component and becomes obvious in panel (f).
}
\end{figure}

Since these measurements are performed with pure inplane field the AHE contribution is weak compared to the much stronger AMR contributions when the data are fit by a serious of $\sin(n\phi)$ terms with $n$ as given in Fig.~\ref{fig:mrrot}d. Figure~\ref{fig:mrrot}d shows the magnetic field dependence of the amplitude of the $\sin(n\phi)$ contributions. By the onset and saturation behavior of the strongest terms with $n=2,4$ the spin-flop field of $\sim 2...3$~T is determined. That a $\sin(3\phi)$ contribution corresponding to AHE signal is nevertheless detected in these magnetic field rotations can be seen when all even contributions are subtracted from the raw data (Fig.~\ref{fig:mr3fold}). What remains is a signal with predominant three-fold symmetry corresponding to the AHE which is shown in Fig.~\ref{fig:mr3fold}b. Consistent with the onset of magnetic moment reorientation at the spin flop field, data measured with fields $>2$~T show a three-fold contribution.

\begin{figure}[tb]
\includegraphics[width=\linewidth]{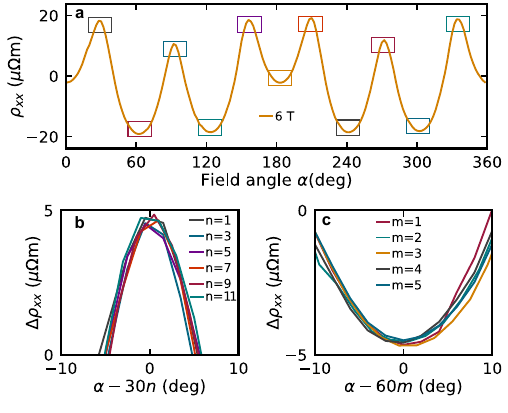}
\caption{\label{fig:wideminima} Angle dependent longitudinal resistivity data recorded at $T=50$~K and magnetic field of 6~T. (a) Longitudinal resistivity variation for an inplane magnetic field rotation with geometry as given in Fig.~\ref{fig:mrrot}. Colored rectangles mark the position and line color of the enlarged view near the maxima (b) and minima (c) of the six-fold magnetoresistance contribution.
}
\end{figure}

In order to determine the magnetic easy axes of the thin films we turn our attention to the longitudinal resistivity data. For high magnetic fields the inplane anisotropy causes the moments to stay closer to the respective easy axis which is nearest to perpendicular to the magnetic field (spin-flop arrangement). Figure~\ref{fig:wideminima} shows examplaric for magnetic field rotations with a field strength of 6 T that the minima of six-fold contribution to the longitudinal resistivity are wider than the respective maxima. These mimina occur for magnetic fields along the $a$-axis and equivalent directions and therefore suggest magnetic easy axes for magnetic moments along $\left<1\bar100\right>$ as shown in Fig.~1b of the main text (or equivalently Fig.~\ref{fig:unitcell}a).

\begin{figure}[tb]
\includegraphics[width=\linewidth]{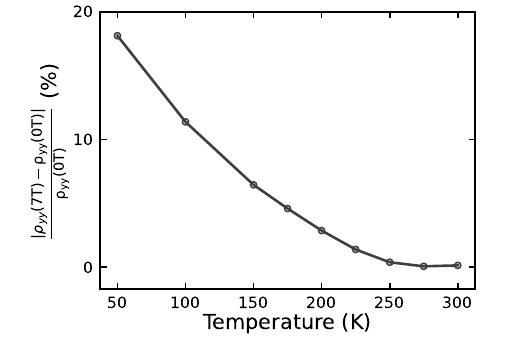}
\caption{\label{fig:mrtemp} Temperature dependence of the magnetoresistance signal determined from the longitudinal resistivity difference between data at 7 and 0 T. The magnetoresistance was extracted for magnetic field sweeps recorded with magnetic field at an angle of $\beta=30^\circ$ and current along $y \parallel \left[01\bar10\right]$.
}
\end{figure}

The temperature dependence of the Hall signal is shown in the main text is furthermore distinct from the magnetoresistance effect which governs the variation of the longitudinal field sweeps which is shown in Fig.~\ref{fig:mrtemp}.

\begin{figure}[tb]
\includegraphics[width=\linewidth]{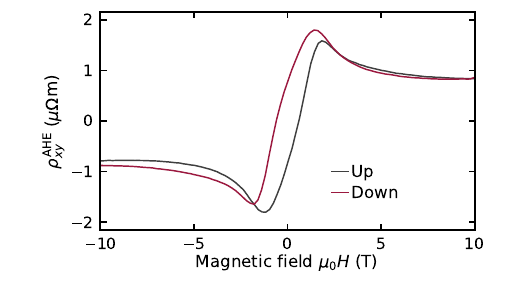}
\caption{\label{fig:rhoxy} Odd in magnetic field transversal resistivity component for current along the $a$-direction ($\left[2\bar1\bar10\right]$) and magnetic field along the current, i.e. inplane. For technical reasons no defined out of plane field component could be set for this current direction.
}
\end{figure}

Since the discussed AHE effect is determined by an out of plane Hall pseudovector the inplane current direction does not matter. We verified this by the use of a Hall bar with current channel along the $x$ direction. The respective data are shown in Fig.~\ref{fig:rhoxy}.

\FloatBarrier

\end{document}